\documentclass{aa}
\begin{document}
\outer\def\gtae {$\buildrel {\lower3pt\hbox{$>$}} \over 
{\lower2pt\hbox{$\sim$}} $}
\outer\def\ltae {$\buildrel {\lower3pt\hbox{$<$}} \over 
{\lower2pt\hbox{$\sim$}} $}
\def\rchi{{${\chi}_{\nu}^{2}$}}
\def\Mdot{\hbox{$\dot M$}}
\newcommand{\dg} {^{\circ}}

\thesaurus{ ()}
\title{Modelling the spin pulse profile of the isolated neutron star RX
J0720.4--3125 observed with {\sl XMM-Newton}\thanks{Based on observations
   obtained with XMM-Newton, an ESA science mission with instruments
   and contributions directly funded by ESA Member States and the USA
   (NASA).}}      
\author{Mark Cropper\inst{1}, Silvia Zane\inst{1}, Gavin
Ramsay\inst{1}, Frank Haberl\inst{2} and Christian Motch\inst{3}} 

\offprints{M. Cropper}

\institute{
Mullard Space Science Laboratory, University College London, Holmbury
St. Mary, Dorking, Surrey, RH5 6NT, UK
\and
Max Planck Institut f\"ur Extraterrestrische Physik,
Giessenbachstrasse, D-85748 Garching, Germany
\and
Observatoire Astronomique, CNRS UMR 7550, 11 Rue de l'Universit\'e,
F-67000 Strasbourg, France
}

\authorrunning{M. Cropper et al.}
\titlerunning{Spin modelling of RX J0720.4--3125}

\date{}

\maketitle

\begin{abstract}

We model the spin pulse intensity and hardness ratio profiles of the isolated
neutron star RXJ0720.4--3125 using {\sl XMM-Newton} data. The observed
variation is
approximately sinusoidal with a peak-to-peak amplitude of 15\%, and the 
hardness
ratio is softest slightly before flux maximum. By using
polar cap models we are able to derive maximum polar cap sizes and 
acceptable viewing
geometries. The inferred sizes of the caps turn out to be 
more compatible with a
scenario in which the neutron star is heated
by accretion, and place limits on the magnetic field strength. The
hardness ratio
modulation can then be explained in terms of energy-dependent beaming
effects, and this constrains the acceptable models of the emerging radiation
to cases in which softer photons are more strongly beamed than harder
photons. An alternative explanation in terms of spatially variable
absorption co-rotating in the magnetosphere is also discussed. 


\end{abstract}

\section{Introduction}
\label{intr} 

From its high ratio of X-ray to optical flux, soft X-ray spectrum, 8.4~s 
X-ray
period and location in the Galactic plane, Haberl et
al.~(\cite{hmb97:1997}) 
proposed that the ROSAT source RXJ0720.4--3125 is an isolated neutron star
(NS). A probable
optical counterpart has been identified (Motch \& Haberl \cite{mh98:1998};
Kulkarni \& van Kerkwijk \cite{kvk98:1998}). Paerels et
al.~(\cite{p00:2001}, Paper~1) present spectra of
RXJ0720.4--3125 using {\it XMM-Newton}, and find that
there is no evidence for
absorption lines and edges in the X-ray spectrum. They are able to refine
the ROSAT spectral fits using black-body spectra of 86.2$\pm$0.3 eV
absorbed
by a cold absorber of $6.0 \pm 0.4 \times 10^{19}$~cm$^{-2}$.

Paper~1 mainly addressed the spectral characteristics and the behaviour of
RXJ0720.4--3125 averaged over the 8.4~s spin period. In this paper we
examine the implications of the phase-resolved data.

\section{Observations}

Observations of RXJ0720.4--3125 were made using {\it XMM-Newton}
(Jansen et al.~\cite{ja01:2001}) on 2000 May 13 (orbit 78). For the phase
resolved data we have concentrated on the EPIC-PN data, as the
EPIC-MOS provides insufficient time resolution in the modes selected
for the observation. The
raw data were processed using the version of the {\sl XMM-Newton} Science
Analysis
System (Watson et al.~\cite{wa00:2001}) released on 2000 July 12. 

\begin{figure}
\begin{center}
\setlength{\unitlength}{1cm}
\begin{picture}(10,8)
\put(-0.5,-3.2){\includegraphics{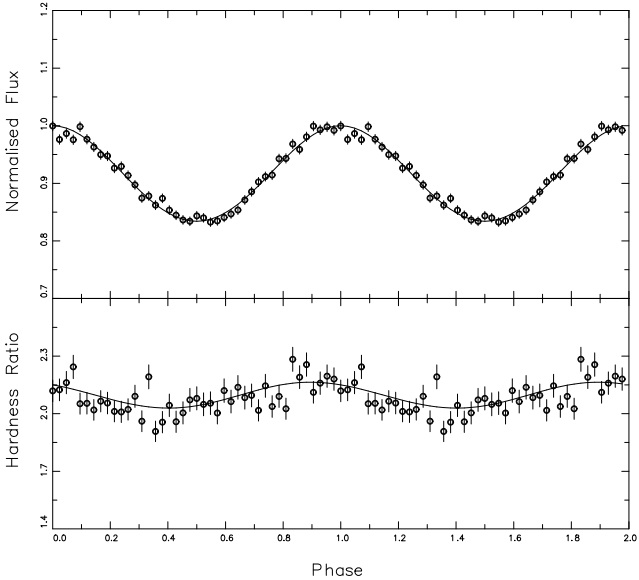}} 
\end{picture}
\end{center}
\caption{
The total flux (top) from EPIC-PN phase averaged on the 8.391~s 
period of Haberl et al.~(\cite{hmb97:1997}) with the (0.1 to 0.4)/(0.4 to
0.8) keV hardness 
ratio (bottom). The flux is normalised to the model flux at phase 0.0. 
Also shown is the best fit to the data
solid polar cap model with beaming and temperature 
variation (see Section \ref{beaming} for details).}
\vspace*{-3mm}
\label{fig:lc} 
\end{figure}

The particle background was significantly higher at the end of the
observation, so those data were removed from the analysis, leaving a
total observing time of 49 ks. Light curves were extracted from the
EPIC-PN camera using an aperture $\sim30^{''}$ in radius centered on RX
0720.4--3125, chosen so that the aperture did not cover more than one
CCD. This radius encompasses $\sim$ 90\% of the integrated PSF
(Aschenbach et al.~\cite{abh00:2000}). We accumulated counts over the energy
ranges
0.1--0.4, 0.4--0.8 and
0.8--1.2 keV and phase-averaged them over 42 phase intervals on the
8.391~s period of Haberl et al.~(\cite{hmb97:1997}). 
The spin profile in the band $0.1-1.2$~keV is shown in
Figure~\ref{fig:lc}, 
together with the hardness ratio (soft/medium) variation and the best fit
discussed in Section \ref{beaming}.

\subsection{Phase--folded X--ray light curves} 

The general shape of the phased intensity curve is similar to that in Haberl et
al.~(\cite{hmb97:1997}) and is described in Paper~1. It appears
approximately symmetrical and sinusoidal with an amplitude $\sim 15$\%.
The hardness ratio is also seen to vary: it is
softest around flux maximum and the amplitude is smaller at $\sim 10$\%.
The
phasing of the hardness ratio curve is slightly but significantly earlier than
the intensity curve. A cross-correlation of the two curves indicates that the
phase difference is $-0.048$.

The most likely source of the flux variation is the changing visibility of the
heated magnetic polar caps (Haberl et al.~\cite{hmb97:1997}) with the
rotation
of the NS. Assuming symmetry, it is possible that the rotation period
is twice that deduced by Haberl et al.~(\cite{hmb97:1997}), depending on
whether one cap or two are seen each rotation. A Fourier analysis excludes
any harmonic or subharmonic to the level of $\sim10$\% of the 8.391~s 
period
but it is possible that odd-even effects may be present at lower levels.  We
have therefore also folded the data on a 16.782~s period. This gives rise
to
a lightcurve with two peaks per spin period with a small difference in
amplitude. In this case the hardness
ratio is also found to show two peaks per spin period.

\begin{figure}
\begin{center}
\setlength{\unitlength}{1cm}
\begin{picture}(10,7.3)
\put(-.8,+7.){\includegraphics{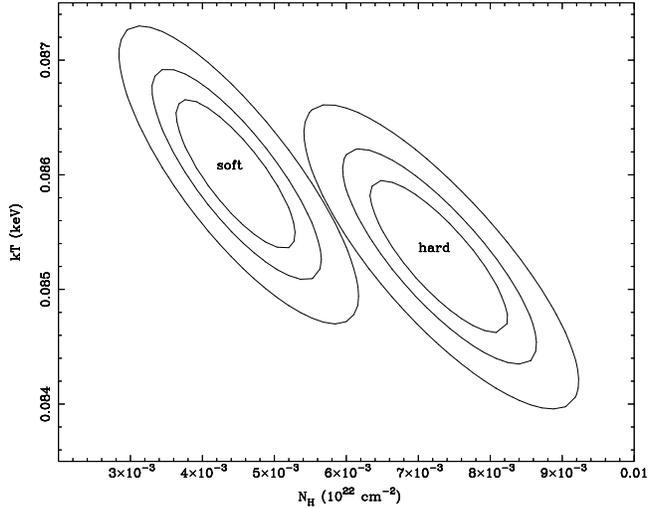}}
\end{picture}
\end{center}
\caption{Confidence contours in the temperature--absorbing column plane
for the phases at the extremes of the hardness ratio variation in Figure
\ref{fig:lc}. Contours are for 68\%, 90\% and 99\% confidence level.
} \label{fig:nh_kt} 
\end{figure}

\subsection{Phase resolved spectra}

We have selected data between phases 0.85--0.05 and 0.15--0.35 to
examine more
closely the change in
spectral parameters in the ``soft'' and ``hard'' phase, i.e. at the
maximum and minimum of the lightcurve, respectively. 
Spectra were background corrected using the same CCD on which the source
was detected. In our fits we used the response file
epn$\_$fs20$\_$sY9$\_$thin.rmf produced by one of us (FH, version 2000 July)
 and we selected only single pixel events. 

As in Paper~1 we have used a simple absorbed black body model in 
XSPEC. The spectral fit suggests that the majority of the variation
is caused by a change in absorption, rather than a change in temperature
(Figure~\ref{fig:nh_kt}). The fits are consistent with a temperature of 85 eV, 
with the measured temperature difference between flux maximum and minimum
of less than 1eV. On the other hand, the best fit of the ``hard'' phase 
requires a  doubling in absorption from $4\times10^{19}$ cm$^{-2}$ to
$8\times10^{19}$ cm$^{-2}$. 
These parameters are consistent with the phase-averaged values
from RGS data derived in Paper~1. While residual uncertainties in the
EPIC-PN
calibration at these soft energies may introduce some uncertainties into the
absolute values of the derived absorptions (see Paper~1), it is unlikely
that the relative
change can be discounted. 
                                                                    
\section{Analysis of the light curve} 

\subsection{Polar cap models} 

The high S/N ratio of the X-ray pulsation in the {\it XMM-Newton} 
data provides the
opportunity for a more detailed modelling than was possible using the 
ROSAT data. In this section, we first present the best fit of the light
curve only, re-adapting the formalism originally derived by 
Pechenick et al.~(\cite{pfc83:1983}). Assuming that the
intensity
variation is due to the changing visibility of heated circular magnetic
polecaps and including gravity effects, those authors were able to derive
a semi-analytical expression for the observed brightness, in terms of
elliptic integrals. Their final result depends on four variables: the
angular radius of the polecap $\alpha$, the angle between the dipole and
rotation axes $\beta$, the angle between the line of sight and the rotation
axis $\gamma$, and $R/2M$, where
$M$ and $R$ are the mass and radius of the NS ($G=c=1$). In
addition, a
prescription for the beaming function, $f(\delta)$, and a phase angle are
required to complete the viewing angle calculation. The equations 
are symmetric for $\beta$ and $\gamma$ and $R/2M$ ranges 
between 2.5 and 4.0 for standard NS masses and equations of state. 

Using this formalism, we have calculated model light curves for a range
of parameter values. We have then used a
Levenberg-Marquardt algorithm to fit the
phased data by least squares. In order to maximise the precision of the
fit we have used the spectral range of 0.1 to 1.2 keV and we have
weighted the fit using the reciprocal of the squares of the uncertainties.

\subsection{Fit to the models} 
\label{lc}

In the case of isotropic emission (no beaming, $f(\delta)=1$) the best fit
to the data occurs for a polar cap radius of $\alpha=35\dg$,
dipole and viewing angles of $\beta=9.3\dg$ and $\gamma=42.7\dg$, and
$R/2M=4.2$. The quality of the fit is good, with \rchi=0.96, and for these
values the cap at the other pole is at most barely visible. The
values of $\alpha$ and $\beta$ are the most sensitive to the angular 
distribution of the radiation field: when we change the beaming 
prescription by adopting a ``pencil'' model ($f(\delta) = \cos \delta$),
the best fit occurs for $\alpha =  12.4$, $\beta =  4.2$, $\gamma = 41.3$
and $R/M = 4.2$ (\rchi=1.14). 

\begin{table}
\begin{center}
\begin{tabular}{crr}
\hline
$f(\delta)$ & $R/2M$  & $\alpha_{max}$  \\
\hline
1             & 4 & $53\dg$   \\
              & 3 & $44\dg$   \\
              & 2.5 & $26\dg$  \\
$\cos \delta$ & 4 & $63\dg$  \\
              & 3 & $61\dg$ \\
              & 2.5 & $59\dg$  \\
$\exp(-\delta)$ & 4 & $65\dg$  \\
                & 3 & $63\dg$   \\
                & 2.5 & $61\dg$  \\       
\end{tabular}
\end{center}
\label{tab1}
\caption{The maximum angular size of the polar caps for different beaming
prescriptions and compactness parameters.} 
\vspace*{-6mm}
\end{table}

\begin{figure}
\begin{center}
\setlength{\unitlength}{1cm}
\begin{picture}(10,7)
\put(-1.8,-3.5){\includegraphics{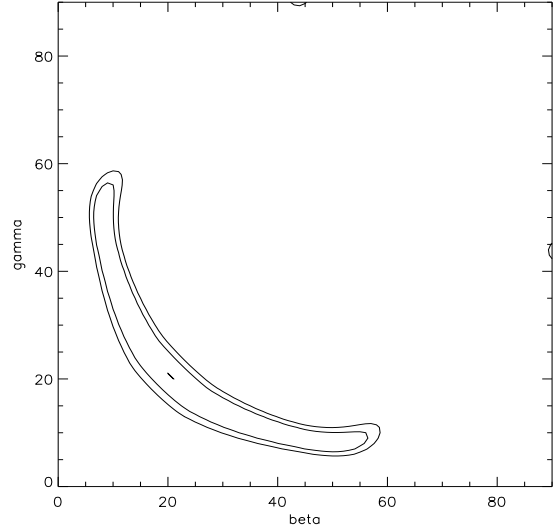}}
\end{picture}
\end{center}
\caption{The \rchi\/ over the $(\beta,\gamma)$ plane for a cap angular 
radius $\alpha=10\dg$ and $R/2M=3.0$. Here $f(\delta)=1$. Contours are for
\rchi=1.0 and then the 68\% and 90\% confidence intervals. The
arc-shaped region and the small
semicircular regions near the axes correspond to periods of 8.391~s and 
16.782~s, respectively.} \label{fig:bg_grid} 
\end{figure}

However, the quasi-sinusoidal nature of the phased intensity data has the
result that the phase space is degenerate for a wide range of input
parameters. We have therefore taken the approach of fixing the cap radius
and $R/2M$ and exploring the $(\beta, \gamma)$ plane for good fits. In the 
case of isotropic emission we have
found that, for most cap radii and for $2.5 \leq R/2M \leq 4.0$, the
acceptable parameter space is concentrated in a hyperbola described
approximately by $\beta\sim C /\gamma$, where $C\sim400$ for angles in
degrees (see figure~\ref{fig:bg_grid}). As the cap size increases the best
fit moves from the center of the arc where $\beta=\gamma$ to
$(\beta,\gamma)\sim(10, 48)$, and the constant $C$ increases slightly.

A physically more important result is that, for a given value of $R/2M$,
the fits to the data constrain the maximum angular radius of the cap. By
increasing $\alpha$, the \rchi~of the best fit in $(\beta, \gamma)$ shows
a sharp increase beyond a threshold value.  We can therefore identify the
maximum polecap radius beyond which a best fit can be rejected at the 90\% 
confidence level (\rchi=4.61). Results are summarized in table~1 for
isotropic
emission, a pencil model and a very extreme case in which the angular
distribution of radiation decays exponentially with $\delta$.  Smaller
$R/2M$ require smaller maximum polecap radii because 
of the increasing bending of the light ray trajectories from stars
with large compactness parameter. In the presence of strong beaming
effects, larger caps are allowed, but, at least when using simple, non
energy--dependent beaming prescriptions, a polecap
larger than $\sim 60\dg -65\dg$ can be rejected at a confidence level of
90\%.

\subsection{Constraints on $B$ from the size of the polar caps}

Independently on the interpretation of RXJ0720.4--3125 as a cooling or an
accreting NS, the existence of an upper limit for
the cap size requires the NS to be magnetised. 

A cooling NS interpretation implies
magnetic fields much larger than in the accreting case. Such a high field,
typically $10^{12}-10^{14}$ G, yields 
temperature variations with the magnetic colatitude,
$\chi$, and modifies the emergent spectrum as function of viewing angle
(Greenstein \& Hartke \cite{gh83:1983}; Page \cite{pa95:1995}). A simple
expression for the surface temperature profile is
(Possenti et al.~\cite{pmc96:1996}):   
\begin{equation} \label{tsu} T_{surf} =
{T_{eff} \over \left [ 1 - 0.47 \left ( 1 - K \right )\right ]^{1/4}}
\left [ { K
+ \left ( 4 - K \right ) \cos^2 \chi \over 1 + 3 \cos^2 \chi } \right
]^{1/4} 
\end{equation} 
where $K = k_\perp/k_{||} $ is the ratio of the coefficients of thermal
conductivity orthogonal and parallel to the field:  $K \sim 10^{-2}-
10^{-4}$ for $B \sim 10^{11}-10^{13}$ G. 
Not even the maximum values of the polar
cap size derived in Section \ref{lc} for strong beaming are 
fully compatible with this kind of interpretation: the
temperature gradient given by equation (\ref{tsu}) is so smooth that, even
in the (unrealistic) limit case $K=0$ the associated blackbody
luminosity drops only by a factor 1/2 for $\alpha\sim 60 \dg$ and by an
order of magnitude for $\alpha \sim 77 \dg$. Both the
emergent spectrum and the degree of modulation of the signal are not
strongly dependent on $K$. 
Despite the real angular distribution of the specific intensity being unknown 
and the fact that a proper modelling of the relative brightness should
include the temperature profile, 
it seems difficult to explain the (relatively) small hot spot
inferred from our calculations in terms of a surface temperature
variation during the NS cooling phase. 

If the polar caps are heated by accretion, the incoming material is
channeled to the poles only by those field lines which have passed the
Alfv\`en radius. In the case of an aligned, dipolar $B$--field this gives
$r_{cap} \sim R^{3/2}/r_A^{1/2}$, where $r_{cap}$, $r_A$ are the cap and
the Alfv\`en radius, respectively (see e.g. equation (9) in Treves et
al.~\cite{ttzc00:2000}). Thus, the previous values of the maximum polecap
radii translate for the case of isotropic emission to
$B>(5\times 10^{5},6\times10^{4},2\times10^{4})\Mdot^{1/2}_{11}$ G, where
$\Mdot_{11}=\Mdot/(10^{11} \rm{g\,s}^{-1})$. 

The minimum polecap radius is
not constrained from the data: we find that good fits can be obtained even
for $\alpha=1\dg$ for all $2.5 \leq R/2M\leq 4.0$. In the
framework of the accreting scenario, the existence of an upper limit $B
\sim 10^{10}$ G can be used to constrain the minimum $\alpha$: for larger
fields, in fact, accretion will be completely hindered. 
This gives a minimum $\alpha \sim 1\dg$, which is indeed consistent with
the data. Such low field strengths may provide one of the first
indication that a significant field decay had occurred in an isolated
NS, or, as it has been also suggested, may indicate that this object is
the outcome of the common envelope evolution of a binary system (see
Treves et al.~\cite{ttzc00:2000} and references therein, Wang
\cite{w97:1997}). 

\section{Simultaneous analysis of light curve and hardness ratio
variation}
\label{beaming} 

The Pechenick et al.~(\cite{pfc83:1983}) formalism assumes emission from a
uniform temperature polecap, and it does not predict any hardness ratio changes
through the spin cycle. However, zones of different temperatures can be used in
conjunction with the spectral response of EPIC-PN and a spectral model for the
emission to calculate the observed fluxes and hardness ratios. We use a
blackbody to simulate the thermal emission as this provides a good
approximation to the flux distribution of more detailed atmospheric models (at
least in the X-ray band, see Zavlin et al.~\cite{zps96:1996}, Zane et 
al.~\cite{ztt00:2000}). However the
angular distribution of the emerging radiation in these models show a variety
of beaming profiles, depending on chemical composition and magnetic field,
which must be accounted for separately in the modelling, as implemented in
Section 3.

The two main observational features are the simultaneous variations of
hardness ratio and luminosity and their shift in phase. In addition, 
the spectral fit at the minimum of the lightcurve implies a larger 
absorption. 
By working on the premise of minimizing the number of additional
assumptions, we investigated the parameter space using different simple
models for the emission. We allow for a temperature
variation inside the caps, for beaming effects, for variations in
absorption and for asymmetric conditions 
in the two spots. Despite the temperature of
the dominant thermal component not varying during a period, the
presence of a
second component with $60 \, {\rm eV} \, <T<  90 \, {\rm eV}$ (or,
equivalently, of a
smooth gradient of $T$) is still acceptable on spectral basis. 

\begin{table}
\scriptsize{
\begin{tabular}{|l|@{}c@{}|@{}c@{}|@{}c@{}|@{}c@{}|@{}c@{}|@{}c@{}|}
\hline
&\multicolumn{3}{c|}{\rule[-1mm]{0mm}{4mm}\bf{8.391~s period}} &\multicolumn{3}{c|}
{\bf{16.782~s
period}}\\
\cline{2-7}
\bf{Model} & \makebox[10.5mm]{2-pole} & \makebox[10.5mm]{2-pole} & \makebox[10mm]{1-pole} & \makebox[10.5mm]{2-pole} & \makebox[10.5mm]{2-pole} & \makebox[10mm]{1-pole} \\
&$\alpha < 35 \dg$ & $\alpha > 35 \dg$
&  & $\alpha < 35 \dg$ & $\alpha > 35 \dg$ &  \\
\hline
\parbox[c]{2.5cm}{2 polar caps at different temperatures}
&\rule[-3mm]{0mm}{7mm} $\times^A$ & $\times^F$ 
& $\times^C$ & $\times^D$ &
$\times^B$ & $\times^B$ \\
\hline
\parbox[c]{2.5cm}{2 polar caps with different absorptions} &
\rule[-3mm]{0mm}{7mm} $\times^A$ & $\times^F$ & $\times^C$ &
$\times^D$ &
$\times^B$ & $\times^B$ \\
\hline
\parbox[c]{2.5cm}{2 equal polar caps with gradient in absorption} &
\rule[-4mm]{0mm}{9mm} $\times^A$
&
{\large $\circ$} &
{\large $\circ$} & {\large$\circ$} & $\times^B$ & $\times^B$ \\
\hline
\parbox[c]{2.5cm}{2 equal polar caps with gradient in temperature}
& \rule[-4mm]{0mm}{9mm} $\times^A$ &
$\times^E$
& $\times^E$ & $\times^E$ & $\times^B$ & $\times^B$ \\
\hline
\parbox[c]{2.5cm}{2 equal polar caps with beaming} & \rule[-3mm]{0mm}{7mm} 
$\times^A$ & {\large $\bullet$} &
{\large $\bullet$} &
{\large $\bullet$} & $\times^B$ & $\times^B$ \\
\hline
\end{tabular}
}
\label{summary}
\caption[]{
Summary of acceptable models:\vspace*{-2mm}
\begin{itemize}
\item[{\large $\bullet$}] Acceptable fits possible.
\item[{\large $\circ$}] Probably unacceptable on grounds of poor model fit and 
unphysicality of absorption model
\item[A] Excluded on the grounds of not being consistent with the 8.391~s  
period.
\item[B] Excluded on the grounds of not being consistent with the 16.782~s
period.
\item[C] Excluded on the grounds that it predicts no hardness ratio variation.
\item[D] Excluded on the grounds that the hardness ratio variation would be on
a 16.782~s period rather than the 8.391~s observed.
\item[E] Excluded on the grounds that the 
temperature difference required to produce the hardness ratio changes is
not consistent with the spectral fit. 
\item[F] Excluded on the grounds of the shape of the hardness ratio variation. 
\end{itemize}
\vspace*{-10mm}}
\end{table}

Our findings are summarized in Table~2. With ``1-pole'' and ``2-pole'' we
refer to models in which one or two polar caps become visible during the
rotation. Quite interestingly, most models
can be discarded since they do not allow adequate fits. This is denoted by an
$\times$ in the table, together with the reason for their exclusion.
We find that limb
darkening variations are the best candidate to give rise to the expected
modulation. 
If the emission is isotropic, the fact that the hardness ratio is {\it
softest} near the
flux maximum can be fitted only if the temperature decreases 
toward the center of the cap; however, the spectral dependence is inconsistent
with the observed spectrum. On the other hand, by
using different prescriptions for $f(\delta)$, we can also allow for a 
more realistic temperature profile, increasing towards the centre.
The observational 
features require a beaming in the softer component much stronger than 
at highest energies. If this is the case, the change 
in $N_H$ found with the simple absorbed blackbody model in XSPEC
should mimic 
artificially the soft excess/deficit at different phases: this is
naturally
predicted by this kind of angular distribution. 
Our best fit model, \rchi=1.23, is shown in Figure 
\ref{fig:lc}.  The fit to the intensity data is good, 
but is not optimised:  because of the number of degrees of
freedom (now including the choice of spectral parameters and phase
offsets) a fit was sought manually to the intensity data which would be
both sufficient in formal terms and predict adequately the hardness ratio
variation in mean value, amplitude and phase. In this model, we allow for
a cooler ring with $T =70$ eV, around a central, hotter  
region of $T = 90$ eV; the phase offset between them is 0.113. 
The value
of the column density has been fixed at $6\times 10^{19}$ cm$^{-2}$, to be
consistent with Paper~1. 
The angular sizes of the soft and hard components are fitted by 
$\alpha_{hot} = 33\dg$, $\alpha_{cold} = 38\dg$, while the other
parameters
are $(\beta,\gamma,R/2M) = (5.3\dg,43.5\dg, 4.2)$. The angular
dependence of the emerging radiation is assumed to change from a 
pencil model, $f(\delta) = \cos \delta $, at low temperatures to isotropic
emission, $f(\delta) = 1$, at high temperatures. No reasonable fit can be
found when the beaming is stronger at high energies, unless we require 
the polar cap to be colder at the center. 

The inferred behaviour of the degree of
anisotropy is contrary to what is physically 
expected at the simplest level in both cooling and accretion atmospheres. 
In an atmosphere in LTE where the main radiative process is 
bremsstrahlung, the opacity decreases with the photon energy. This, in 
turn, implies the anisotropy in the angular distribution being smaller at
low frequencies (since these photons escape from the outermost
atmospheric layers) and increases with the energy when photons decouple
in the deepest, hottest layers. Moreover, as pointed out by Zavlin
et al.~(\cite{zps96:1996}) for pure H cooling models, the
maximum of the intensity spectrum shifts towards lower energies with
increasing $\delta$, contrary to the hardness ratio variations. 
In order to match the hardness ratio variation, the energy--dependence of
the degree of anisotropy must be reversed. For hydrogen atmospheres, that
may happen only in the very high energy part of the spectrum, were the 
angular dependence tends to match the Ambartsumian--Chandrasekhar function 
for scattering dominated propagation (see Figure 8 in Zavlin et
al.~\cite{zps96:1996}). 

The observed change in $N_H$ may be real (see Paper~1), but, at least
within our
simple assumptions, models with different levels
of absorption do not produce acceptable fits. Moreover, a variation in
absorption over the spin cycle require the absorbing material to be close
to the NS, probably co-rotating in
the magnetosphere. In this case the light cylinder limits the
maximum distance to the absorber to $R_{max}\sim4\times10^{10}$ cm. On
the other hand, in order for the material to be substantially neutral, the
ionisation parameter $\xi=L/nR^2$ must be less than 1. This in turn 
requires a particle density $n>10^{10}$ cm$^{-3}$, for $R<R_{max}$ and a
luminosity of $10^{31}$ erg s$^{-1}$ (Haberl et al.~\cite{hmb97:1997}).
A similar result can be
obtained
from the inferred increase in absorption of $4\times10^{19}$ cm$^{-2}$.
For this to occur within $R_{max}$ then $n>10^9$ cm$^{-3}$. 
For comparison, typical densities of the interstellar medium are $n\sim 1$
cm$^{-3}$. The implication is that a reservoir of accreting material is
held in the magnetosphere, in regions where gravitational, centrifugal and
magnetic forces balance and confine it. These densities may be considered
uncomfortably high, although we are unable to exclude them on current
knowledge. We return to this in Section 5.

\section{Conclusions} 

By exploiting the superior quality of the data obtained from
{\it XMM-Newton \/}, we have been able to present 
a detailed analysis of the spin pulse profile of 
RXJ0720.4--3125. 
We have modelled these data by using polar cap models based on
the Pechenick et al.~(\cite{pfc83:1983}) formulation. The
smooth nature of the data does not allow a unique conclusion
about the best fit values to be reached, but nevertheless we are able
to exclude a large volume of the parameter space and to derive
an upper limit for the size of the emitting region. The derived values are
at best marginally compatible with an interpretation in terms of
surface temperature variation during the cooling phase of a middle-age NS. An
accretor model is more satisfactory in accounting for the intensity plus
hardness-ratio variations. Accreting NS appear to be relatively rare objects,
as recently found by Popov et al.~\cite{pct00:2000} from a statistical
evolutionary analysis. RXJ0720.4--3125 may therefore be more important than
perhaps initially appreciated.

We have investigated a number of possible explanations for the
intensity and spectral variations over the spin period. 
The long period of RXJ0720.4--3125 is somewhat unusual and hints at a possible
evolutionary link between dim NS, soft gamma-ray repeaters (SGRs) and anomalous
X-ray pulsars (AXPs) (Haberl et al.~\cite{hmb97:1997}, Alpar~\cite{a00:2000}), 
whereby isolated NS
in the propellor phase are the progenitors of AXPs and SGRs. If the hardness
ratio variation is indeed caused by absorption, then the large amount of
material close by to the neutron star may be explained in terms of matter being
propelled outwards. However the luminosity in this phase is caused by energy
dissipation in the neutron star, and we have already found (Section 3.3) this
requires polar
cap sizes which are uncomfortably large. Moreover, such an evolutionary link
would produce significantly larger numbers of AXPs and SGRs (Mereghetti, 
private communication).

The most plausible
interpretation is that the observed modulation is produced by beaming
effects in the emerging radiation. 
The angular distribution of the emerging radiation is 
still a matter for further investigation, and is 
extremely sensitive to details in the radiative computations and to a 
number of assumptions about chemical composition and magnetic field
(see e.g. Zavlin et al.~\cite{zps96:1996}), Rajagopal et
al.~\cite{rrm97:1997}). Our modelling provides observational inputs for
these investigations by suggesting that the softer X-rays from the heated
cap are more strongly beamed than the harder X-rays.

\begin{acknowledgements}

We are grateful to R.~Turolla and A.~Possenti for several very
helpful discussions, and to the referee for constructive comments.

\end{acknowledgements}

\end{document}